\documentstyle[12pt]{article}

\begin{document}
\begin{titlepage}
\title{\begin{flushright} {\normalsize HEP-Th  9403xxx \\ February
1994 } \\
\end{flushright}  \bigskip   \bigskip
{\bf  The zero-curvature representation \\ for
     nonlinear O(3) sigma-model }}
\author{\\ {\bf A. G. Bytsko}
\thanks{ e-mail: bytsko@lomi.spb.su $\;\;\;$ and $\;\;\;$
bytsko@diada.spb.su }
 \\
 Steklov Mathematical Institute \\
 Fontanka 27, St.Petersburg 191011 \\ Russia     \\  }
\date{\it{Dedicated to Professor L.D. Faddeev \\ on the occasion of
his
sixtieth birthday \\}}
\maketitle \thispagestyle{empty}
 \begin{abstract}
 {\rm  We consider $O(3)$ sigma-model as a reduction of the principal
chiral
 field. This approach allows to introduce the currents with
ultralocal
 Poisson brackets and to obtain the zero-curvature equation which
 admits the fundamental Poisson bracket.}
\end{abstract}
\end{titlepage}

\section*{ Introduction }

 In present paper we consider one of the field theory models - the
nonlinear $O(3)$ sigma-model in the framework of the inverse
scattering method.
The important step in this approach is to find a zero-curvature
representation
for the investigated model. An attempt to obtain this representation
for
$O(3)$ sigma-model was done in paper [1]. But in this paper some
extra
conditions on components of energy-momentum tensor were imposed.
Besides,
essential points in [1] were the transition to Sine-Gordon type
variables
and application of the Backlund transformation. This classically
admittable
transition is a serious obstacle for using the quantum version of the
inverse
scattering method.

 On the other hand, one may consider $O(3)$ sigma-model as a
reduction of a
more general model - that of the principal chiral field. The
zero-curvature
representation for the principal chiral field was obtained in paper
[2]. The
quantum inverse scattering method was applied to this model in paper
[3].
An essential trick in [3] was the change of Poisson brackets for the
currents.
In present paper we realize this scheme for $O(3)$ sigma-model by
means of
construction of new pair of currents. But, as we shall see, it is
impossible
to get the XXX-magnetic model in a way similar to that in the paper
[3].

 Thus, in present paper new currents are constructed for the
investigated
model and new $U-V$ pair is obtained via these variables. Also the
fundamental
Poisson bracket is written out and some properties of new currents
are
obtained.

\section{ $\vec{n}$-field model }

 The nonlinear $O(3)$ sigma-model ( it is also known as
$\vec{n}$-field model
 ) describes a three-dimensional vector $\vec{n}=(n_{1},n_{2},n_{3})$
with components depending on coordinates in (1+1) dimensional
space-time. The
vector $\vec{n}$ satisfies the space-periodisity condition :
    \[ \vec{n}(x+L,t)=\vec{n}(x,t)   \]

 The Lagrangian of the model coincides with Lagrangian of the free
field :
  \begin{equation}  \label{1}
     {\cal L}  =\frac{1}{2} \int_{0}^{L} ((\partial_{0}\vec{n})^2   -
   (\partial_{x}\vec{n})^2) dx
\end{equation}
(here we denote : $ \partial_{0} \equiv \frac{\partial}{\partial t}$
;
 $ \partial_{x} \equiv \frac{\partial}{\partial x}$ ).
 We also suppose that the vector $\vec{n}$ satisfies an extra
condition -
its length is fixed :
  \begin{equation}  \label{2}
     \vec{n}^2 = n_{1}^{2} + n_{2}^{2} + n_{3}^{2} = 1
\end{equation}
 The equations of motion following from Lagrangian (\ref{1}) and
respecting
constraint (\ref{2}) are :
 \begin{equation}  \label{3}
  \partial_{0}^2 \vec{n} - \partial_{x}^2
\vec{n}+((\partial_{0}\vec{n})^2
 - (\partial_{x}\vec{n})^2 )\vec{n} = 0
\end{equation}
 For Hamiltonian description  of the model we introduce the density
of
momentum :
\[ \vec{\pi} \equiv \frac{\partial {\cal
L}}{\partial(\partial_{0}\vec{n})} =
 \partial_{0}\vec{n}  \]
The Legendre transformation of the Lagrangian (\ref{1}) gives the
Hamiltonian
of the model :
 \[ H = \int_{0}^{L} (\vec{\pi} \cdot \partial_{0}\vec{n}) dx - {\cal
L} =
   \frac{1}{2} \int_{0}^{L} (\vec{\pi}^{2} +
(\partial_{x}\vec{n})^{2})dx  \]
 The Poisson brackets of variables $\vec{\pi}$ and $\vec{n}$ can be
obtained
from the canonical brackets as Poisson-Dirac brackets with respect to
constraints
 \[       \vec{n}^2 = 1 \;\;\;\;\; , \;\;\;\;\;\;\;\;\;\;\; \\
  \vec{\pi}\cdot\vec{n} = 0           \]
These brackets are :
 \begin{equation}  \label{4}
 \begin{array}{l}
   \{ n^{a}(x),n^{b}(y)\} =0 \;\;\;\;\; ,
\;\;\;\;\;\;\;\;\;\;\;\;\;\;\;

\{{\pi}^{a}(x),n^{b}(y)\}=(\delta^{ab}-n^{a}(x)n^{b}(x))\delta(x-y)
\\
         \\
\{ {\pi}^{a}(x),{\pi}^{b}(y)\}= ({\pi}^{b}(x)n^{a}(x) -
{\pi}^{a}(x)n^{b}(x))\;
   \delta(x-y)      \\
   \end{array}
\end{equation}
Because the model is $O(3)$-invariant, it is more efficient to
describe it in
terms of variables $\vec{l}$ and $\vec{n}$ , where $\vec{l}$ (the
angular
momentum) is defined as :
\[   \vec{l}(x) = \vec{\pi}(x) \wedge \vec{n}(x) \;\;\;\;, \\
 \;\;\;\;   \vec{l}\;^2(x) = \vec{\pi}^2(x)      \]

 Here $\wedge$ denotes a vector product :
$l^{a} = \epsilon^{abc} {\pi}^{b} n^{c}$ .

The Poisson brackets of these variables
 \begin{equation}  \label{5}
  \{ l^{a}(x),l^{b}(y) \} = \epsilon^{abc} l^{c}(x) \delta(x-y)
\end{equation}
 \begin{equation}  \label{6}
  \{ l^{a}(x),n^{b}(y) \} = \epsilon^{abc} n^{c}(x) \delta(x-y)
\end{equation}
 \begin{equation}  \label{7}
  \{ n^{a}(x),n^{b}(y) \} = 0
\end{equation}
define the current algebra of group $E(3)$. The phase space of the
model is a
simplectic orbit :
 \begin{equation}  \label{8}
           \vec{n}^2 = 1 \;\;\;\; , \;\;\; \vec{l} \cdot \vec{n} = 0
\end{equation}

\section{ Standard zero-curvature representation }

 The basic method for investigation of models corresponding to
different nonlinear equations is the inverse scattering method.
Nowadays this
is a well developed tool to study classical as well as quantum
versions of
models (see for example [4]). This method starts from representation
of the
nonlinear equation in the form of the zero-curvature condition :
 \begin{equation}  \label{9}
  \partial_{t} U(x,\lambda) - \partial_{x} V(x,\lambda) +
[U(x,\lambda),
  V(x,\lambda)] = 0
\end{equation}

Here $U$ and $V$ are the same size square matrices with elements
depending on
space-time variables $x$ and $t$ as well as on some extra spectral
parameter
$\lambda$. Equations (\ref{9}) have to be satisfied for all values of
$\lambda$.

 Let us consider the principal chiral field model. In this model the
dynamical
variable $g(x,t)$ takes the values in some compact group $G$.
Equations of
motion have the following form :
  \begin{equation}  \label{10}
    \partial_{0}^2 g - \partial_{x}^2 g = \partial_{0}g \cdot g^{-1}
\cdot
\partial_{0}g - \partial_{x}g \cdot g^{-1} \cdot \partial_{x}g
 \end{equation}

 An appropriate $U-V$ pair for this model was found by V.Zakharov and
A.Mikhailov [2] in form :
  \begin{equation}  \label{11}
  U(x,\lambda) = \frac{1}{2} \frac{L_{0}(x) + L_{1}(x)}{1-\lambda} -
\\
  \frac{1}{2} \frac{L_{0}(x) - L_{1}(x)}{1+\lambda} =
\\
   \frac{\lambda L_{0}(x) + L_{1}(x)}{1-{\lambda}^2}
 \end{equation}
 \begin{equation}  \label{12}
  V(x,\lambda) = \frac{1}{2} \frac{L_{0}(x) + L_{1}(x)}{1-\lambda} +
\\
  \frac{1}{2} \frac{L_{0}(x) - L_{1}(x)}{1+\lambda} =
\\
   \frac{\lambda L_{1}(x) + L_{0}(x)}{1-{\lambda}^2}
 \end{equation}

Here the left currents are introduced, taking the values in Lie
algebra of
group $G$ :
   \begin{equation}  \label{13}
 L_{0}(x,t)= \partial_{0}g \cdot g^{-1} \;\;,\;\;
  L_{1}(x,t)= \partial_{x}g \cdot g^{-1}
\end{equation}
The zero-curvature condition (\ref{9}) for $U-V$ pair
(\ref{11}),(\ref{12})
can be written via these currents $L_{\mu}$ :
 \begin{equation}  \label{14}
 \partial_{0}L_{0} - \partial_{x}L_{1} = 0
 \end{equation}
 \begin{equation}  \label{15}
 \partial_{0}L_{1} - \partial_{x}L_{0} + [L_{1},L_{0}] = 0
 \end{equation}

Let us note that the equation (\ref{15}) follows from definitions
(\ref{13}):
\[ \partial_{0}(\partial_{x}g \cdot g^{-1})
-\partial_{x}(\partial_{0}g
\cdot g^{-1}) \\
+ [\partial_{x}g \cdot g^{-1},\partial_{0}g \cdot g^{-1}] \equiv 0 \]
while equation (\ref{14}) is a consequence of equations of motion
(\ref{10}).

The Lagrangian and the Hamiltonian can be expressed using the
currents
$L_{\mu}$ :
 \begin{equation}  \label{16}
{\cal L} = \frac{1}{4} \int_{0}^{L} tr(L_{1}^2 - L_{0}^2)dx \;\;\; ;
\;\;\; \\
   H = - \frac{1}{4} \int_{0}^{L} tr(L_{0}^2 + L_{1}^2)dx
 \end{equation}

Poisson brackets of components of currents $L_{\mu}^{a}$ are :
 \begin{equation}  \label{17}
 \{ L_{0}^{a}(x), L_{0}^{b}(y) \} = \epsilon^{abc} L_{0}^{c}(x)
\delta (x-y)
 \end{equation}
 \begin{equation}  \label{18}
\{ L_{0}^{a}(x), L_{1}^{b}(y) \} = \epsilon^{abc} L_{1}^{c}(x) \delta
(x-y) \\
 - \delta^{ab} \delta^{\prime}(x-y)
 \end{equation}
 \begin{equation}  \label{19}
 \{ L_{1}^{a}(x), L_{1}^{b}(y) \} = 0
 \end{equation}

In general case of the principal chiral field with values in the
group $G$
the equations of motion (\ref{10}) admit an extra reduction :
 \begin{equation}  \label{20}
                       g^2 (x,t) = I
 \end{equation}
This reduction does not change the form of $U-V$ pair, zero-curvature
equations (\ref{14}),(\ref{15}) and relations (\ref{16}). It also
preserves
the Poisson brackets (\ref{17})-(\ref{19}) because brackets of the
currents
with the constraint (\ref{20}) vanish.

 Applying the reduction (\ref{20}) to the case of the principal
chiral field
for $G = SU(2)$, we obtain the $O(3)$ sigma-model. Indeed, in this
case the
variable $g(x,t)$ can be written in form :
 \[ g(x,t) = \vec{n} \cdot \vec{\sigma} \equiv n^{a}(x,t)
{\sigma}_{a}
  \;\;\;\; , \;\; \vec{n}^2 = 1 \]
where $\sigma^{a}$ are Pauli matrices :
\[ \begin{array}{ccc}
 {\sigma}_{1} =  \left(  \begin{array}{cc}  0  & 1 \\
                                          1 & 0    \end{array}
\right) \; ,
  &  {\sigma}_{2} =  \left(  \begin{array}{cc}  0  & -i \\
                                          i & 0    \end{array}
\right) \; ,
  &  {\sigma}_{3} =  \left(  \begin{array}{cc}  1  & 0 \\
                                          0 & -1    \end{array}
\right) \;
\end{array}   \]
For given parametrization of $g(x,t)$ equations of motion (\ref{10})
turn
into equations of motion (\ref{3}) for $\vec{n}$-field.

  Using a relation
 \[ (\vec{a}\vec{\sigma})\cdot (\vec{b}\vec{\sigma})=(\vec{a}
\vec{b})
 \cdot I + i(\vec{a} \wedge \vec{b})\cdot \vec{\sigma}  \]
we can express the currents $L_{\mu}= \sum_{a=1}^{3} i
L_{\mu}^{a}\sigma_{a}$
via variables of $\vec{n}$-field :
 \begin{equation}  \label{21}
  L_{0}(x,t)= i(\partial_{0} \vec{n} \wedge \\
  \vec{n}) \cdot \vec{\sigma} \equiv i \vec{\l} \cdot \vec{\sigma}
 \end{equation}
 \begin{equation}  \label{22}
  L_{1}(x,t)= i(\partial_{x} \vec{n} \wedge  \vec{n}) \cdot
\vec{\sigma}
 \end{equation}

Thus, if we consider $O(3)$ sigma-model as a reduction of the
principal chiral
field for group $G = SU(2)$, we immediately obtain a zero-curvature
representation for it with $U-V$ pair given by (\ref{11})-(\ref{12})
and
with the equations for currents (\ref{14})-(\ref{15}). As we have
mentioned
above, one of these equations is a consequence of the equations of
motion
(\ref{3}) :
 \[  \partial_{0}L_{0} - \partial_{x}L_{1} = i(\partial_{0}^2 \vec{n}
 \wedge \vec{n} - \partial_{x}^2 \vec{n} \wedge\vec{n}) \cdot
\vec{\sigma}
  = i((\partial_{0}^2 \vec{n} - \partial_{x}^2 \vec{n})\wedge\vec{n})
  \cdot \vec{\sigma} = 0 \]

To apply further the inverse scattering method for solving the
$\vec{n}$-field model one should represent the Poisson brackets of
elements
of matrix $U(x,\lambda)$ as a fundamental Poisson bracket :
 \begin{equation}  \label{23}
  \{\; \stackrel{1}{U}(x,\lambda)\;,\;\stackrel{2}{U}(y,\mu)\;\} =
\\
  {[}\; r (\lambda , \mu ) \; ,\; \stackrel{1}{U} (x,\lambda) \; +
  \stackrel{2}{U} (x,\mu) \;{]} \; \delta(x-y)
 \end{equation}

Here $r(\lambda)$ is some classical $r$-matrix and we use the
notations :
\[ \begin{array}{cc}
 \stackrel{1}{U} = U \otimes I \;\;\;   ,  & \stackrel{2}{U} = I
\otimes U
  \end{array}  \]

Unfortunately, since the formula (\ref{18}) has a term with
$\delta^{\prime}(x-y)$ (so-called nonultralocal term), it is
impossible to
introduce an appropriate fundamental Poisson bracket.

\section{ Currents $J_{\mu}$ and new $U-V$ pair }

 In order to find the new $U-V$ pair which satisfies a zero-curvature
representation and has an ultralocal fundamental Poisson bracket we
define
new current variables :
 \begin{equation}  \label{24}
  J_{0}(x,t) = \frac{1}{2} i (\partial_{0} \vec{n} \wedge \vec{n} - i
  \partial_{x} \vec{n}) \cdot \vec{\sigma} \equiv \frac{1}{2} i
(\vec{l} - i
  \partial_{x} \vec{n}) \cdot \vec{\sigma}
 \end{equation}
 \begin{equation}  \label{25}
  J_{1}(x,t) = \frac{1}{2} i (\partial_{x} \vec{n} \wedge \vec{n} - i
  \partial_{0} \vec{n}) \cdot \vec{\sigma} \equiv \frac{1}{2} i
(\partial_{x}
   \vec{n} \wedge \vec{n} - i \vec{\pi} ) \cdot \vec{\sigma}
 \end{equation}

 It should be noticed here that in case of the group $SU(2)$ the
addition of terms like $\frac{1}{2} \partial_{\mu}g= \frac{1}{2}
\partial_{\mu}\vec{n} \cdot \vec{\sigma}$ to the currents $L_{\mu}$
does not
bring them out of Lie algebra $su(2)$. Consequently, $J_{\mu}$ may be
regarded
indeed as current variables. But if we apply the formulas given above
for
construction of new currents in case of arbitrary Lie group $G$, the
new
objects $J_{\mu}$ will not belong to Lie algebra of this group (but,
nevertheless, they will satisfy the zero-curvature equations).

 Let us consider an analogue of the pair (\ref{11})-(\ref{12}) for
currents
$J_{\mu}$ :
 \begin{equation}  \label{26}
  U^{\prime}(x,\lambda) = \frac{ \lambda J_{0}(x) + J_{1}(x) }{ 1 -
  {\lambda}^{2} } \;\;\; , \;\;\;
   V^{\prime}(x,\lambda) = \frac{ \lambda J_{1}(x) + J_{0}(x) }{ 1 -
   {\lambda}^{2} }
 \end{equation}

The zero-curvature condition (\ref{9}) leads to the system of
equations :
 \begin{equation}  \label{27}
  \partial_{0} J_{0} - \partial_{x} J_{1} = 0
 \end{equation}
 \begin{equation}  \label{28}
  \partial_{0} J_{1} - \partial_{x} J_{0} = [ J_{0},J_{1} ]
 \end{equation}

Formally, the system (\ref{27})-(\ref{28}) coincides with the system
(\ref{14})-(\ref{15}), but it should be noticed that now both
equations
(\ref{27}) and (\ref{28}) follow from the equations of motion
(\ref{3}) :
  \[ \partial_{0} J_{0} - \partial_{x} J_{1} = \frac{1}{2} i
  ( \partial_{0}^{2} \vec{n} \wedge \vec{n} - i \partial_{0x} \vec{n}
-                \\
\partial_{x}^{2}  \vec{n} \wedge \vec{n} + i \partial_{x0} \vec{n} )
\\
 \cdot \vec{\sigma} = \\
\frac{1}{2} i (( \partial_{0}^{2} \vec{n} -  \partial_{x}^{2}
\vec{n})
\wedge \vec{n}) \cdot \vec{\sigma} = 0   \]

 In order to check the equation (\ref{28}) one has to write down the
new
currents $J_{\mu}$ in the form :
\[ J_{0} = \frac{1}{2} \partial_{0}g \cdot g^{-1} + \frac{1}{2}
\partial_{x}g
 = \frac{1}{2} ( L_{0} + L_{1} g)  \]
\[ J_{1} = \frac{1}{2} \partial_{x}g \cdot g^{-1} + \frac{1}{2}
\partial_{0}g
 = \frac{1}{2} ( L_{1} + L_{0} g)  \]
and use the condition of reduction
 \begin{equation}  \label{29}
  g^{2} = (\vec{n} \cdot \vec{\sigma})^{2} = \vec{n}^{2} \cdot I
\equiv I
  \;\;\;\;\;\;\; ,
 \end{equation}
as well as its consequences
 \begin{equation}  \label{30}
  g^{-1} = g
 \end{equation}
and
 \begin{equation}  \label{31}
   L_{\mu}\; g = - g L_{\mu}
 \end{equation}
(The identity (\ref{31}) was obtained by differentiation of the
equation
(\ref{30})) \footnote{ It is interesting to note, that the relation
(\ref{31})
leads to coincidence of right and left currents in this model :
 \[ R_{\mu} \equiv - g^{-1} \cdot \partial_{\mu}g = \partial_{\mu}g
\cdot
 g^{-1} \equiv L_{\mu} \;\;\; . \]  }

Now l.h.s. of equation (\ref{28}) may be written in the form
  \[  \partial_{0} J_{1} - \partial_{x} J_{0} = \frac{1}{2}
(\partial_{0} L_{1}    \\
   - \partial_{x} L_{0}) + \frac{1}{2} (\partial_{0} L_{0} -
\partial_{x} L_{1})g +   \\
  \frac{1}{2} (L_{0}^{2} - L_{1}^{2} )g \;\;\; , \]

and taking into account the equations (\ref{14})-(\ref{15}) (in fact
this
means we use the equations of motion) we come to relation
\[ \partial_{0} J_{1} - \partial_{x} J_{0} = \frac{1}{2}
[L_{0},L_{1}]  + \\
\frac{1}{2} (L_{0}^{2} - L_{1}^{2} )g  \]

The r.h.s. of equation (\ref{28}) may be obtained in the same form
using
the conditions (\ref{29})-(\ref{31}) :
\[ [J_{0},J_{1}] = \frac{1}{4} [ L_{0} + L_{1}g , \; L_{1} + L_{0}g ]
\\
  = \frac{1}{4}  [L_{0},L_{1}] + \frac{1}{4} [L_{0},L_{0}g]  +
\frac{1}{4} [L_{1}g,L_{1}]+
    \frac{1}{4} [L_{1}g,L_{0}g ] =                   \]
\[ = \frac{1}{4} [L_{0},L_{1}] +  \frac{1}{4} (L_{0}^{2}g -
L_{0}gL_{0}) + \\
 \frac{1}{4}(L_{1}gL_{1}- L_{1}^{2}g) + \frac{1}{4}(L_{1}gL_{0}g -
L_{0}gL_{1}g)= \]
\[ = \frac{1}{4} [L_{0},L_{1}] + \frac{1}{2}L_{0}^{2}g -
\frac{1}{2}L_{1}^{2}g + \frac{1}{4} [L_{0},L_{1}]g^{2} \\
 = \frac{1}{2} [L_{0},L_{1}] + \frac{1}{2}(L_{0}^{2} - L_{1}^{2})g
  \]

Thus, the difference from case of unreduced principal chiral field is
that
the second $U^{\prime}-V^{\prime}$ pair for $\vec{n}$-field exists.
The next
step on the way of the considered model investigation is the
calculation of
Poisson brackets for new currents.

\section{ Fundamental Poisson bracket }

 Now let $t^a$, a = 1,2,3 be the basic elements of the algebra
$su(2)$. They
are supposed to be normalized with respect to the Killing form :
 \[ tr(t^{a}t^{b}) = - \frac{1}{2} \delta^{ab}\;\;\; ; \;\;
[t^{a},t^{b}] =
 f^{abc}t^{c}   \]

 The generators $t^a$ may be expressed via Pauli matrices :
\[   t^{a}= \frac{1}{2}i{\sigma}^{a}\;\;\; ; \;\;\; f^{abc} =
- {\epsilon}^{abc}  \]

In order to calculate the Poisson brackets of components of the
current
$J_{0}(x)=J_{0}^{a}(x)t^{a}$ we use the formulas (\ref{5})-(\ref{7})
\footnote{ We need also a well known property of generalized
functions :
 \[ (xF(x),\delta^{\prime} (x)) = -(F(x),\delta (x)) \;\;\;\;\;\; .
\] } :

  \begin{equation}  \label{32}
 \begin{array}{c}
 \{J_{0}^{a}(x),J_{0}^{b}(y)\} = \{l^{a}(x) - i\partial_{x} n^{a}(x),
 l^{b}(y) - i\partial_{y} n^{b}(y)\} = \\
     \\
= \epsilon^{abc}l^{c}(x) \delta (x-y) +
i\epsilon^{abc}(n^{c}(x)-n^{c}(y))\; \delta^{\prime} (x-y) = \\
\\
 = \epsilon^{abc}l^{c}(x) \delta (x-y) +i\epsilon^{abc}((x-y)
\partial_{x} n^{c}(x))\; \delta^{\prime} (x-y) = \\
  \\
= \epsilon^{abc}l^{c}(x) \delta (x-y) - i\epsilon^{abc}\partial_{x}
n^{c}(x)\; \delta (x-y) =
 \epsilon^{abc} J_{0}^{c}(x)\; \delta (x-y)
 \end{array}
 \end{equation}

 We have also the following relations :
  \begin{equation}  \label{33}
 \{J_{0}^{a}(x),n^{b}(y) \} = \{l^{a}(x) - i\partial_{x}
n^{a}(x),n^{b}(y)\}
  = \epsilon^{abc} n^{c}(x) \; \delta (x-y)
 \end{equation}
 \begin{equation}  \label{34}
 \{ J_{1}^{a}(x),n^{b}(y) \} = \{ (\partial_{x} \vec{n} \wedge
\vec{n} -
 i\vec{\pi})^{a}(x) \;,\; n^{b}(y) \} = i( n^{a}(x)n^{b}(x)-
\delta^{ab})\;
  \delta(x-y)
 \end{equation}

In order to obtain the other Poisson brackets it is convenient to use
the
following relations for currents (let us recall that
$\vec{l}\cdot\vec{n}=
\vec{\pi}\cdot\vec{n}= \partial_{x}{\vec{n}} \cdot\vec{n}= 0 $ ) :
 \begin{equation}  \label{35}
  i\;(\vec{J}_{1} \wedge \vec{n} ) = i\;( \partial_{x}\vec{n}\wedge
\vec{n}
  -i \vec{\pi})\wedge \vec{n}
  = \vec{\pi}\wedge \vec{n} -i\partial_{x}\vec{n} = \vec{J}_{0}
 \end{equation}
 \begin{equation}  \label{36}
   i\;(\vec{J}_{0} \wedge \vec{n}) = i\;(\vec{l}
-i\partial_{x}\vec{n})\wedge
   \vec{n}
  = \partial_{x}\vec{n}\wedge \vec{n} -i \vec{\pi} = \vec{J}_{1}
 \end{equation}

(Here we denote : $\vec{J}_{\mu}\;=\;(J^{1}_{\mu},J^{2}_{\mu},
 J^{3}_{\mu})$ ).

 Taking into account the relation $\vec{J}_{\mu} \wedge
  \vec{n} = - \vec{n} \wedge \vec{J}_{\mu} \;$,  one may rewrite the
formulas
(\ref{35})-(\ref{36}) in the following way :
  \begin{equation}  \label{37}
   J_{1}\;g = - g\;J_{1} = J_{0}
 \end{equation}
  \begin{equation}  \label{38}
  J_{0}\;g = - g\;J_{0} = J_{1}
 \end{equation}

Now, using the formulas (\ref{32})-(\ref{36}) we obtain
  \begin{equation}  \label{39}
    \begin{array}{c}
 \{ J^{a}_{0}(x)\;,J^{b}_{1}(y) \}  = i\epsilon^{bcd}\;\{
J^{a}_{0}(x)\;,
 J^{c}_{0}(y)n^{d}(y)\} =\\
 \\
 = i\epsilon^{bcd}\;(\epsilon^{acf}\;J^{f}_{0}(x)n^{d}(y) +
 \epsilon^{adf}\;J^{c}_{0}(y)n^{f}(x))\;\delta(x-y) =\\
 \\
 = i\;(\; - J^{b}_{0}(x)n^{a}(x) +
J^{a}_{0}(x)n^{b}(x))\;\delta(x-y)=
 \epsilon^{abc}\;J^{c}_{1}(x)\;\delta(x-y)
\end{array}
 \end{equation}

Similarly one can find :
   \begin{equation}  \label{40}
\{ J^{a}_{1}(x)\;,J^{b}_{1}(y) \} =
\epsilon^{abc}\;J^{c}_{0}(x)\;\delta(x-y)
 \end{equation}

 An important property of brackets (\ref{32}),(\ref{39}),(\ref{40})
is their
ultralocality (they do not include terms like
$\delta^{\prime}(x-y)$). This
allows us to write down a fundamental Poisson bracket for matrix
$U^{\prime}(x,\lambda)$ in $\vec{n}$-field model.

 Indeed, using (\ref{32}),(\ref{39}),(\ref{40}) and following [4] we
find
 \[ \{\; \stackrel{1}{U^{\prime}}(x,\lambda)\;,\;
 \stackrel{2}{U^{\prime}}(y,\mu)\;\} = \\
 -\frac{1}{4} \frac{\sigma^{a}\otimes \sigma^{b}}{(1
 -\lambda^{2})(1-\mu^{2})}\\
 \{\; \lambda J^{a}_{0}(x) + J^{a}_{1}(x)\; , \;
 \mu J^{b}_{0}(y)+J^{b}_{1}(y)\;\}=\]
 \[=-\frac{1}{4} \frac{\epsilon^{abc}\sigma^{a}\otimes
 \sigma^{b}}{(1-\lambda^{2})(1-\mu^{2})}\\
\;((\lambda \mu + 1)J^{c}_{0}(x) + (\lambda +
\mu)J^{c}_{1}(x))\;\delta(x-y) \]

 Now, introducing the permutation operator
 \[ P = \frac{1}{2}(I \otimes I + \sigma^{a}\otimes \sigma^{a})  \]
and using its property
\[ \epsilon^{abc} \sigma^{a}\otimes \sigma^{b} =
-i\;{[}\;P\;,\;\sigma^{c}
\otimes I \;{]}  = -2 {[} \; P \;,\;\ t^{c} \otimes I \; {]} \;\;\; ,
  \]
we obtain
\[ \{\; \stackrel{1}{U^{\prime}}(x,\lambda) \; ,
\;\stackrel{2}{U^{\prime}}
(y,\mu)\;\} = \\
\frac{{[} \;P\;,\;(\lambda\mu +1)\stackrel{1}{J_{0}}(x) +
(\lambda+\mu) \\
 \stackrel{1}{J_{1}}(x)\; {]} }{2(1-\lambda^{2})(1-\mu^{2})}\;
\delta(x-y)= \]

\[=\frac{1}{(1-\lambda^{2})(1-\mu^{2})} {[}\;\frac{P}{2(\lambda-\mu)}
\;,\;
(\lambda^{2}-1)\mu \stackrel{1}{J_{0}}(x) \\
  + (1-\mu^{2})\lambda\stackrel{1}{J_{0}}(x) + (\lambda^{2}-1)
  \stackrel{1}{J_{1}}(x) + \]
 \[ +(1-\mu^{2}) \stackrel{1}{J_{1}}(x) {]} \; \delta(x-y)=  \\
{[}\;\frac{P}{2(\lambda-\mu)}\;,\;
\stackrel{1}{U^{\prime}}(x,\lambda) \; -
  \stackrel{1}{U^{\prime}}(x,\mu)\;{]} \; \delta(x-y) =\]
 \[={[}\;\frac{P}{2(\lambda-\mu)}\;,\;
\stackrel{1}{U^{\prime}}(x,\lambda) \\
 \; + \stackrel{2}{U^{\prime}}(x,\mu)\;{]} \; \delta(x-y) \]

So, we have found the fundamental Poisson bracket for matrix
$U^{\prime}(x,\lambda)$ with the following $r$-matrix :
\[ r(\lambda,\mu) = \frac{P}{2(\lambda-\mu)} \;\;\;\;\;  \]

 We have succeeded to find the fundamental Poisson brackets due to
the
observation that the transition from standard currents $L_{0}$ and
$L_{1}$
(which correspond by formulas (\ref{21})-(\ref{22}) to the pair of
variables
$\vec{l}$ and $\partial_{x} \vec{n} \wedge \vec{n}$) to currents
$J_{0}$ and
$J_{1}$ leads to disappearence of the term with
$\delta^{\prime}(x-y)$ in
Poisson brackets. At first sight, however, this disappearence
looks quite unexpected. It may be explained by the fact that new
currents are not deformations of standard currents.

 Indeed, if one takes the initial Hamiltonian with an extra parameter
$\gamma$ :
 \[ H = \frac{1}{2} \int_{0}^{L} (\; \vec{\pi}^2 + {\gamma}^2
(\partial_{x}
 \vec{n})^2 \;)\;dx \;\;\;, \]
he will obtain the following standard and new currents :
 \[ \begin{array}{ll}
  \vec{L}_{0}(x) = \vec{l}(x) \;\;\;\;\;{;}\;\;\;\;\;\;
\vec{L}_{1}(x) =
  \gamma \partial_{x} \vec{n} \wedge \vec{n}  \\
   \\
  \vec{J}_{0}(x) = \vec{l}(x) -\gamma \partial_{x} \vec{n} \;\;\;
{;}\;\;\;
 \vec{J}_{1}(x) = \gamma \partial_{x}\vec{n} \wedge \vec{n} - i
\vec{\pi} =
  -i ( \vec{\pi} + i\gamma \partial_{x}\vec{n} \wedge \vec{n} )
  \end{array}  \]

These formulas imply that one can consider the currents $J_{0}$ and
$J_{1}$
as deformations with parameter $\gamma$ of variables $\vec{l}$ and
$-i \vec{\pi}$. One can come to the same conclusion simply comparing
the
relations (\ref{32}),(\ref{39}),(\ref{40}) for $J_{0}$ and $J_{1}$
with
formulas (\ref{4})-(\ref{5}) for $\vec{l}$, $\vec{\pi}$.

\section{ Lagrangian, Hamiltonian and currents $\hat{J}_{\mu}$ }

 As was mentioned above, the Lagrangian as well as the Hamiltonian of
$\vec{n}$-field may be written via currents $L_{\mu}$ :
 \begin{equation}  \label{41}
   {\cal L} = \frac{1}{4} \int_{0}^{L} tr(L_{1}^2 - L_{0}^2) \; dx
\\
   = \frac{1}{2} \int_{0}^{L} (\vec{l}^2 -(\partial_{x} \vec{n}
\wedge \\
   \vec{n})^2) \; dx
 \end{equation}
 \begin{equation}  \label{42}
   H = - \frac{1}{4} \int_{0}^{L} tr(L_{0}^2 + L_{1}^2) \; dx
   = \frac{1}{2} \int_{0}^{L} (\vec{l}^2 +(\partial_{x} \vec{n}
\wedge \\
   \vec{n})^2) \; dx
 \end{equation}

It turns out that the analogous formulas can be obtained in terms of
new
currents $J_{\mu}$. Indeed, if we consider the Lagrangian :
 \[ {\cal L}^{\prime} = \frac{1}{2} \int_{0}^{L} tr(J_{1}^2 -
 J_{0}^2)dx =\\
  \int_{0}^{L} tr \; J_{1}^2 \; dx =  - \int_{0}^{L} tr \; J_{0}^2 \;
dx = \]
 \begin{equation}  \label{43}
  = \frac{1}{2} \int_{0}^{L} (\vec{l} -i\partial_{x} \vec{n})^{2} \;
dx \\
= \frac{1}{2} \int_{0}^{L} (\vec{l}^{2} - (\partial_{x} \vec{n})^{2})
\; dx \\
   - i \int_{0}^{L} (\vec{l} \cdot \partial_{x}\vec{n})\; dx
 \end{equation}
it will differ from the initial Lagrangian (\ref{1}). But the
difference of
these Lagrangians
 \[ \Theta = {\cal L}^{\prime} - {\cal L} = i \int_{0}^{L}\\
  (\vec{l} \cdot \partial_{x}\vec{n})\; dx =\\
 i \int_{0}^{L} (\partial_{x}\vec{n} \wedge \partial_{0}\vec{n})
\cdot
 \vec{n} \; dx     \]
coincides with so-called topological term which does not give a
contribution
into the variation of the action (see for example [5]). Indeed :
 \[ S_{\Theta} = \int_{t_{1}}^{t_{2}} \Theta \; dt \\
  = i\int_{t_{1}}^{t_{2}} \\
  (d\vec{n} \wedge d\vec{n}) \vec{n} \equiv i\int_{t_{1}}^{t_{2}}
\Omega \]
and, since the form $\Omega$ is closed $ d \Omega = 0$ , it leads to
$\delta S = i\int_{t_{1}}^{t_{2}} d \Omega = 0$ . Thus, the
Lagrangians
${\cal L}^{\prime}$ and ${\cal L}$ are equivalent.

 The Hamiltonian of the model may be written in form :
 \[ H= - \int_{0}^{L} tr J_{0} \stackrel{\wedge}{J_{0}} dx =\\
  \int_{0}^{L} tr J_{1}\stackrel{\wedge}{J_{1}} dx = \\
 \frac{1}{2} \int_{0}^{L} (\vec{l}^2 +(\partial_{x} \vec{n})^2) dx
\]

Here we introduce one more pair of currents which up to the sign are
Hermitian conjugations of currents $J_{\mu}$ :
 \begin{equation}  \label{44}
 \stackrel{\wedge}{J_{0}}(x,t) \equiv - J_{0}^{*}(x,t)= \frac{1}{2}\\
  i (\vec{l} + i \partial_{x} \vec{n})  \cdot \vec{\sigma}
 \end{equation}
 \begin{equation}  \label{45}
 \stackrel{\wedge}{J_{1}}(x,t) \equiv - J_{1}^{*}(x,t)=   \\
  \frac{1}{2} i (\partial_{x} \vec{n} \wedge \vec{n} + i \\
  \vec{\pi} ) \cdot \vec{\sigma}
 \end{equation}

 One may easily check that the currents $\stackrel{\wedge}J_{\mu}$
satisfy
the zero-curvature equations (\ref{27})-(\ref{28}), while equations
(\ref{37})-(\ref{38}) turn into the equations :
 \begin{equation}  \label{46}
  g\; \stackrel{\wedge}{J_{1}} = - \stackrel{\wedge}{J_{1}}\;g =\\
  \stackrel{\wedge}{J_{0}}
 \end{equation}
 \begin{equation}  \label{47}
   g\; \stackrel{\wedge}{J_{0}} = - \stackrel{\wedge}{J_{0}}\;g =\\
    \stackrel{\wedge}{J_{1}}
 \end{equation}

The Poisson brackets of currents $\stackrel{\wedge}J_{\mu}$ exactly
repeat
the formulas (\ref{32}),(\ref{39}),(\ref{40}) :
   \begin{equation}  \label{48}
 \begin{array}{c}
\{
\stackrel{\wedge}{J^{a}_{0}}(x)\;,\;\stackrel{\wedge}{J^{b}_{0}}(y)
\}
 = \epsilon^{abc}\;\stackrel{\wedge}{J^{c}_{0}}(x)\;\delta(x-y) \\
 \{
\stackrel{\wedge}{J^{a}_{0}}(x)\;,\;\stackrel{\wedge}{J^{b}_{1}}(y)
\}
 = \epsilon^{abc}\;\stackrel{\wedge}{J^{c}_{1}}(x)\;\delta(x-y) \\
 \{
\stackrel{\wedge}{J^{a}_{1}}(x)\;,\;\stackrel{\wedge}{J^{b}_{1}}(y)
\}
 = \epsilon^{abc}\;\stackrel{\wedge}{J^{c}_{0}}(x)\;\delta(x-y) \\
  \end{array}
 \end{equation}

But, as one should expect, the Poisson brackets of currents $J_{\mu}$
with
$\stackrel{\wedge}J_{\mu}$ contain nonultralocal terms :
\[ \{J^{a}_{0}(x)\;,\;\stackrel{\wedge}{J^{b}_{0}}(y) \} =\\
\epsilon^{abc}\;l^{c}(x)\;\delta(x-y) - i\epsilon^{abc} \\
(n^{c}(x) + n^{c}(y))\; \delta^{\prime}(x-y) =   \]
\[ = \epsilon^{abc}\;J_{0}^{c}(x)\;\delta(x-y) - 2i\epsilon^{abc}
n^{c}(x) \\
\delta^{\prime}(x-y) \]

 \[ \{J^{a}_{0}(x)\;,\;\stackrel{\wedge}{J^{b}_{1}}(y) \} =\\
\epsilon^{abc} \stackrel{\wedge}{J^{c}_{1}}(x) \; \delta(x-y)  \\
 - 2 \delta^{ab} \; \delta^{\prime}(x-y)  \]

  \[ \{ \stackrel{\wedge}{J^{a}_{0}}(x)\;,\;J^{b}_{1}(y) \} =\\
\epsilon^{abc} J^{c}_{1}(x) \; \delta(x-y)  \\
 - 2 \delta^{ab} \delta^{\prime}(x-y)  \]

One should notice that the presence of currents
$\stackrel{\wedge}J_{\mu}$
together with $J_{\mu}$ in the Hamiltonian is quite natural. Indeed,
one can obtain four new variables from standard currents $L_{\mu}$ :
\[ J_{0} = \frac{1}{2} ( L_{0} + L_{1}g ) \;\;\;\; , \;\;\;\; \\
   J_{1} = \frac{1}{2} ( L_{1} + L_{0}g )   \]
\[ \stackrel{\wedge}{J_{0}} = \frac{1}{2} ( L_{0} - L_{1}g ) \;\;\;\;
,\\
 \;\;\;\; \stackrel{\wedge}{J_{1}} = \frac{1}{2} ( L_{1} - L_{0}g )
\]

As we can conclude from relations (\ref{37})-(\ref{38}) and
(\ref{46})-
(\ref{47}) the pair $J_{0}$, $J_{1}$ as well as the pair
$\stackrel{\wedge}J_{0}$, $\stackrel{\wedge}J_{1}$ are not, in fact,
independent variables. So, in general case, in order to keep the
number of
independent variables one has to use, for example, the pair of
currents
$J_{0}$, $\stackrel{\wedge}J_{0}$. And from this point of view, it is
rather
unexpected that formulas for $U-V$ pair and expression for Lagrangian
contain,
in fact, only one independent current.

 Here we may give as illustrations the following expressions for
$U-V$ pair :
\[ U^{\prime}(\lambda) = \frac{ \lambda J_{0}(x) + J_{1}(x) }{ 1
- {\lambda}^{2} } = J_{0} (g + \lambda) [ (g - \lambda) (g + \lambda)
]^{-1} =
J_{0} (g - \lambda)^{-1} = - (g + \lambda)^{-1} J_{0} \]
\[ V^{\prime}(\lambda) = \frac{ \lambda J_{1}(x) + J_{0}(x) }{ 1
- {\lambda}^{2} } = J_{1} (g - \lambda)^{-1} = - (g + \lambda)^{-1}
J_{1} \]

 Finally, we note that if we replace in formulas (\ref{26}) the
currents
$J_{\mu}$ with the currents $\stackrel{\wedge}J_{\mu}$, it will give
us one
more pair of matrices $U^{\prime \prime}(x,\lambda)$,
$V^{\prime \prime}(x,\lambda)$ satisfying the zero-curvature
equation. It
follows from the relation (\ref{48}) that the matrix
$U^{\prime \prime}(x,\lambda)$ will have the same fundamental Poisson
bracket
and $r$-matrix as the matrix $U^{\prime}(x,\lambda)$ has.

\section{ Light-cone coordinates }

Since the Lagrangian (\ref{1}) is relativistically invariant, it is
natural to
use light-cone coordinates
\[   \xi = \frac{t + x}{2} \;\;\;\;\;,\;\;\;\;\; \eta = \frac{t -
x}{2} \]

It is convenient to introduce simultaneously new variables :
\[ J_{+}(x,t) = \sum_{a=1}^{3} J_{+}^{a} t^{a} = \frac{1}{2}
( J_{0}(x,t) + J_{1}(x,t) )\\
= \frac{1}{4} \partial_{\xi} g \cdot g^{-1} + \frac{1}{4}
\partial_{\xi} g \]
\[ J_{-}(x,t) = \sum_{a=1}^{3} J_{-}^{a} t^{a} = \frac{1}{2}
( J_{0}(x,t) - J_{1}(x,t) )\\
= \frac{1}{4} \partial_{\eta} g \cdot g^{-1} - \frac{1}{4}
\partial_{\eta} g\]

It follows from (\ref{37})-(\ref{38}) that
 \begin{equation} \label{49}
 J_{+} \; g = - g \; J_{+} = J_{+}
 \end{equation}
\begin{equation} \label{50}
  g \; J_{-} = - J_{-} \; g = J_{-}
 \end{equation}

Using (\ref{32}),(\ref{39}),(\ref{40}) we get :
 \begin{equation} \label{51}
 \{J_{+}^{a}(x)\;,\;J_{+}^{b}(y) \} = \epsilon^{abc} J_{+}^{c}(x) \\
 \;  \delta (x-y)
 \end{equation}
 \begin{equation} \label{52}
 \{J_{-}^{a}(x)\;,\;J_{-}^{b}(y) \} = \epsilon^{abc} J_{-}^{c}(x) \\
 \;  \delta (x-y)
  \end{equation}
 \begin{equation} \label{53}
 \{J_{+}^{a}(x)\;,\;J_{-}^{b}(y) \} = 0
  \end{equation}

Now, following the paper [3], we may consider the variables $S$ and
$T$ as a
pair of independent spin variables. In [3] the transition from
principal
chiral field model to XXX-magnetic model was performed on this way,
and the
well-developed approach of the Bethe ansatz was used for further
investigation.

But in our case this scheme cannot be applied directly. Indeed, one
can obtain from (\ref{35})-(\ref{38}) the following relations :
\begin{equation} \label{54}
    J_{0}^{2} = - J_{1}^{2}
  \end{equation}
\begin{equation} \label{55}
  J_{0}\;J_{1} = - J_{0} \; g \; J_{0} = - J_{1}\;J_{0}
  \end{equation}

 From these relations we immediately conclude :
\begin{equation} \label{56}
 S^{2} \equiv J_{+}^{2} = \frac{1}{4}(J_{0} + J_{1})^{2} = 0
\;\;\;,\;\;\; \\
 T^{2} \equiv J_{-}^{2} = \frac{1}{4}(J_{0} - J_{1})^{2} = 0
  \end{equation}

 Thus, the Casimir operators corresponding to the spin variables
$\vec{S}$ and
 $\vec{T}$ have zero values. This fact does not allow us to use the
Bethe
 ansatz approach as it was done in [3].

 Let us note that relation (\ref{56}) gives some extra condition when
solving an auxiliary system of differential equations. In our case
taking into account the concrete form of matrices
$U^{\prime}(x,\lambda)$
and $V^{\prime}(x,\lambda)$ (\ref{26}) and using light-cone
coordinates,
we can write down the auxiliary problem as follows
\begin{equation} \label{57}
 \partial_{\xi} \vec{F} = \frac{2}{1 - \lambda} J_{+} \vec{F}
\;\;\;\; ,
 \;\;\;\; \partial_{\eta} \vec{F} = \frac{2}{1 + \lambda} J_{-}
\vec{F}
  \end{equation}

 The relations (\ref{56}) lead to the extra conditions :
\begin{equation} \label{58}
  J_{+} \; \partial_{\xi} \vec{F} = 0 \;\;\;\; , \;\;\;\;
  J_{-} \; \partial_{\eta} \vec{F} = 0
  \end{equation}

Using the relations (\ref{49})-(\ref{50}), one can rewrite the same
conditions
in the form :
 \[ (g + 1) \partial_{\xi} \vec{F} = 0 \;\;\;\; , \;\;\;\;
   (g - 1) \partial_{\eta} \vec{F} = 0  \]

Finally, let us note that the relations (\ref{57}),(\ref{58}) allow
to write
down the second order differential equations for vector $\vec{F}$ :
\begin{equation} \label{59}
\partial_{\xi}^{2} \vec{F} = \frac{2}{1-\lambda}(\partial_{\xi}J_{+})
\vec{F}
 \;\;\;\; , \;\;\;\; \partial_{\eta}^{2} \vec{F} =
\frac{2}{1+\lambda}
 (\partial_{\eta}J_{-}) \vec{F}
  \end{equation}

\section{ Conclusion }

Thus, as it was shown above, the fact that obtained currents
$\vec{J}_{+}$,
$\vec{J}_{-}$ have the zero lengths does not allow to apply a
standard
scheme of quantization. But the same fact points out another possible
approach for the investigation of the model - that of the transition
to
fermion-type variables. Indeed, if we write down the currents
$\vec{J}_{+}$,
$\vec{J}_{-}$ as follows
\[ \begin{array}{cc}
 \vec{J_{+}}   = \frac{1}{2} \left(  \begin{array}{c}  i (z_1^2 -
 \stackrel{-}{z}_1^2) \\ z_1^2 \;+ \stackrel{-}{z}_1^2  \\
 2 i z_1  \stackrel{-}{z}_1         \end{array} \right) \;\; , \;\;
  \vec{J_{-}}   = \frac{1}{2} \left(  \begin{array}{c}  i (z_2^2 -
  \stackrel{-}{z}_2^2) \\ z_2^2 \;+ \stackrel{-}{z}_2^2  \\
 2 i z_2  \stackrel{-}{z}_2        \end{array} \right)
\end{array}   \]

where $z_{1}(x)$ and $z_{2}(x)$ are the components of the spinor
$\Psi(x)$
with the Poisson brackets :
\[ \{ z_{i}(x) \;,\; \stackrel{-}{z}_{j}(y) \} =
\delta_{ij}\;\delta(x-y) \]
we will easily check the formulas (\ref{51})-(\ref{53}) and the
relation :
\begin{equation} \label{56}
   (\vec{J_{+}})^2 = (\vec{J_{-}})^2 = 0
  \end{equation}

Thus, in principle one is able to quantize the $\vec{n}$-field model
by means
of the fermionization of current variables. We are going to perform a
more
detailed investigation in the forthcoming paper.

\section*{ Acknowledgements }

 I am very grateful to L.D. Faddeev for permanent supervision. I
would like
 to thank M.A. Semenov-Tian-Shansky for useful discussions.


\section*{ References }

1. K. Pohlmeyer. Integrable Hamiltonian Systems and Interactions
through
Quadratic Constraints .- Comm.Math.Phys. v.46 No.3 (1976), pp.207-221

2. V.E. Zakharov, A.V. Mikhailov.- Sov.Phys. JETP (Engl.Transl) 74
(1978),
p.1017

3. L.D. Faddeev, N.Yu. Reshetikhin. Integrability of the Principal
Chiral
Field Model in 1+1 Dimension.- Annals of Physics, 167 (1986), pp.227
- 256

4. L.D. Faddeev, L.A. Takhtajan. Hamiltonian methods
in the theory of solitons. Springer-Verlag, 1986.

5. L.D. Faddeev. In quest of the multidimensional solitons. In book:
Nonlocal, nonlinear and nonrenormalizable field theories. (IV
Conference
on nonlocal field theories. USSR, Alushta, 1976) (In Russian)
pp.207-223
(Engl. Transl. to be published in Advanced Series in Math.Phys.)

\end{document}